\documentclass[amsmath,superscriptaddress,prb,twocolumn] {revtex4}
\usepackage{bm}
\usepackage{enumerate}
\usepackage{xcolor}

\usepackage{graphicx}
\usepackage[dvips]{epsfig}
\usepackage{epsf}

\makeatletter
\newcommand{\Rmnum}[1]{\expandafter\@slowromancap\romannumeral #1@}
\makeatletter

\begin{document}

\title{Anomalous Hall effect in metallic collinear antiferromagnets}

\author{Vladimir P. Golubinskii}
\affiliation{National University of Science and Technology MISIS, Moscow, 119049 Russia }

\author{Vladimir A. Zyuzin}
\affiliation{L.D. Landau Institute for Theoretical Physics, 142432, Chernogolovka, Russia}

\begin{abstract}
We propose and theoretically study minimal models of N\'{e}el ordered collinear (compensated) antiferromagnets that show the anomalous Hall effect. For simplicity, we first consider two-dimensional models of antiferromagnets with two magnetic sublattices on a square lattice. We provide explicit examples of a N\'{e}el ordered ferrimagnet and a Dzyaloshinskii weak ferromagnet. Then antiferromagnet on the rutile lattice, that belongs to the class of weak ferromagnets, is studied.
We analyze Dzyaloshinskii's invariants for the existence of spontaneous magnetization in these N\'{e}el ordered systems. Microscopic calculations of the Berry curvature for the studied systems confirm the validity of these Dzyaloshinskii's invariants. It is shown that the anomalous Hall effect mechanism in these antiferromagnets arises either from the interplay of momentum-dependent exchange interaction of conducting fermions with the N\'{e}el order and the spin-orbit coupling. These physical processes originate from the broken symmetries that permit the Dzyaloshinskii's invariant in the system.
\end{abstract}
\maketitle

\section{Introduction}

Metallic collinear N\'{e}el ordered antiferromagnets can exhibit the anomalous Hall effect (AHE) despite having a seemingly vanishing net magnetic moment. There are two types of N\'{e}el ordered antiferromagnets that show the AHE \cite{Turov1965}. For the sake of the argument let's keep in mind a collinear N\'{e}el order with two magnetic sublattices.
The first type are the Dzyaloshinskii weak ferromagnets \cite{BorovikRomanov1957,Dzyaloshinskii1958}. In such antiferromagnets, the two magnetic sublattices are connected by some combination of crystal lattice symmetry operations and the time-reversal operation.
Weak ferromagnetism suggests that a finite magnetic moment can arise in collinear N\'{e}el ordered antiferromagnets due to spin-orbit coupling (SOC), provided the lattice symmetry and the direction of the N\'{e}el vector permit it. All weak ferromagnets for all symmetry classes were classified in \cite{Turov1965}, and their properties were studied and reviewed in \cite{VonsovskiiTurov1986,Turov1990}. 
It is important to note that the magnetic moment may be realized either in the canting of the N\'{e}el order\cite{BorovikRomanov1957,Dzyaloshinskii1958}, or it may result in the orbital magnetization of conducting fermions that interact with the collinear N\'{e}el order\cite{TurovShavrov,VonsovskiiTurov1986,Turov1990}.
The second type are the ferrimagnets, where the two magnetic sublattices are not connected by any combination of crystal lattice symmetry operations and time-reversal operation. For instance, a ferrimagnet can be realized on a lattice where the two magnetic sublattices, possessing equal-magnitude spins, have different non-magnetic environments.

There are also genuine collinear N\'{e}el ordered antiferromagnets in which crystal symmetry forbids the existence of a finite magnetic moment of the Dzyaloshinskii weak ferromagnetism type. Such systems may possess a symmetry that involves a combination of translation and time-reversal operation, or $\pi/2$ rotation and time-reversal operation, or mirror reflection and time-reversal operation, which connects the two magnetic sublattices.

Weak ferromagnetism has been theoretically studied in \cite{Turov1965,VonsovskiiTurov1986,Turov1990, Moriya1960b,Shekhtman1992,Solovyev1997}, and the AHE in them was considered in \cite{TurovShavrov,VonsovskiiTurov1986,Turov1990,AHE_AFM,Solovyev2025}. Different models of ferrimagnets in relation to AHE have been recently theoretically studied in \cite{Naka2020,Zyuzin2024,Zyuzin2025a,Zyuzin2025b}. 

In addition to classifying antiferromagnets based on the existence or absence of a finite magnetic moment, there is currently an ongoing research interest in understanding the spin-splitting of conducting fermions that interact with the N\'{e}el order \cite{Noda2016,Okugawa2018,HayamiYanagiKusunose2019,Ahn2019,AHE_AFM,Rashba2020,HayamiYanagiKusunose2020,Naka2019,Brekke2023,AgterbergPRB2024,Zyuzin2024,Zyuzin2025a}.
For example, one can distinguish $d$-, $g$-, $i$-, or mirror-symmetric type spin splittings based on which symmetry operation connects the Fermi surfaces of opposite spins of conducting fermions. This symmetry corresponds to the way the magnetic sublattices are connected to each other \cite{Turov1965}.
Such momentum-dependent spin splittings can be found in any of the discussed types of antiferromagnets: genuine antiferromagnets, ferrimagnets, or weak ferromagnets.
As a result of their symmetries, genuine antiferromagnets can show unusual effects \cite{VonsovskiiTurov1986,Turov1990} like quadratic and $d$-wave symmetric Faraday rotation in a magnetic field \cite{KharchenkoBibikEremenko1985,EremenkoKharchenko1987}, $d$-wave Hall effect and linear magnetoconductivity \cite{VonsovskiiTurov1986,Turov1990,VorobevZyuzin2024}, and the relevant for spintronics spin-splitter effect \cite{Naka2019} and spin anomalous Hall effect \cite{Zyuzin2025a}. Transport and optical properties \cite{Solovyev1997} of weak ferromagnets have been studied and reviewed in Ref.~\onlinecite{VonsovskiiTurov1986,Turov1990}. 

Despite of the significant research into the AHE in collinear N\'{e}el ordered antiferromagnets Ref.~\onlinecite{TurovShavrov,VonsovskiiTurov1986,Turov1990,AHE_AFM,Naka2020,Zyuzin2025a,Solovyev2025}, the underlying mechanism remains not fully understood and discussed.
The purpose of this paper is to theoretically understand microscopic details of the AHE in collinear N\'{e}el ordered antiferromagnets, introduce Dzyaloshinskii's invariant, and present simple examples of all three types (genuine, ferrimagnets, and weak ferromagnets) of antiferromagnets.

%-----------------------------------------------------------------------------------------------------------------
\section{Dzyaloshinskii's invariants}
%-----------------------------------------------------------------------------------------------------------------
We will consider N\'{e}el order collinear antiferromagnet with two magnetic sublattices with ${\bf M}_{1/2} =\pm {\bf m}$ magnetization. For the analysis of the existence of the magnetic moment in the system, it is necessary to introduce a N\'{e}el vector ${\bf L} = {\bf M}_{1}-{\bf M}_{2}$ and the magnetization ${\bf M} = {\bf M}_{1}+{\bf M}_{2}$.  In principle, ${\bf M}$ can be general, not necessary related to ${\bf M}_{1}+{\bf M}_{2}$.

For example, genuine antiferromagnets are those which always have ${\bf M} = 0$, while ferrimagnets and weak ferromagnets allow for ${\bf M}\neq 0$. In the analysis of existence of magnetic moment, we first set the N\'{e}el order on some lattice and assume that ${\bf M} = 0$. Then we study a question whether ${\bf L}$ can generate a finite ${\bf M}$ in the system. Theoretically, it is a question of whether a $M_{\alpha}L_{\beta}$ combination (in general odd in ${\bf L}$), which may appear in the free energy of the system, is invariant under all symmetries of the crystal. Such a term in free energy is the source term, in which the N\'{e}el vector $L_{\beta}$ is the generator of finite magnetization $M_{\alpha}$ in the system. 
Turov \cite{Turov1965} has classified such invariants in antiferromagnets for all crystal systems. We will refer to such invariants as Dzyaloshinskii's invariants.

In this paper we first study two-dimensional antiferromagnets, since they allow for transparency of the analysis and offer rather simple analytics. The approach is then applied to the antiferromagnet on a rutile lattice. 
In two-dimensions ($x-y$ plane is the plane of the system) in an antiferromagnet with two magnetic sublattices we are expecting N\'{e}el order generated magnetic moment ${\bf M}$ to be normal to the plane of the system ($z-$direction). Below we will be interested in the case when ${\bf M}$ isn't related to canting of the N\'{e}el order, i.e. to ${\bf M}_{1}+{\bf M}_{2}$.
When performing symmetry analysis, we must remember that both ${\bf M}$ and ${\bf L}$ change under the symmetry operations as pseudovectors. In addition, ${\bf L}$ changes sign when the magnetic sublattices are exchanged.

%-----------------------------------------------------------------------------------------------------------------
\section{Berry curvature}
%-----------------------------------------------------------------------------------------------------------------
We analyze the Berry curvature for various models of metallic N\'{e}el ordered antiferromagnets.
The Berry curvature defines the intrinsic mechanism of the AHE. Furthermore, the Berry curvature probes the finite orbital magnetization $\mathbf{M}$ carried by conducting fermions \cite{AHE_RMP,BerryReview,Vanderbilt2018} that interact with the N\'{e}el order. 
Berry curvature $\Omega^{(\pm)}_{{\bf k};\alpha\beta}$ for a general $2 \times 2$ Hamiltonian in the spin space 
\begin{align}
\hat{H}_{\mathrm{eff}}= \left[
\begin{array}{cc} 
\delta_{\bf k}  & \chi_{\bf k}^{*} \\
\chi_{\bf k} & -\delta_{\bf k}
\end{array}
\right],
\end{align}
where $\delta_{\bf k}$ is real (unitary matrix in the Hamiltonian doesn't define the Berry curvature),
is
\begin{align}
&
 \Omega_{{\bf k};\alpha\beta;\pm} = \pm \frac{w_{{\bf k};\alpha\beta} }{2 (\delta_{\bf k}^2 + \vert \chi_{\bf k} \vert^2 )^{\frac{3}{2}}},
\end{align}
where the index $\pm$ is related to the eigenvalues $\epsilon_{\mathbf{k};\pm} = \pm \sqrt{\delta_{\mathbf{k}}^2 + \vert \chi_{\mathbf{k}}\vert^2}$, while $\alpha$ and $\beta$ define projections of the momentum $\mathbf{k}$. We have introduced the function
\begin{align}\label{omega}
w_{{\bf k};\alpha\beta}&=  \delta_{\bf k}
\left[ 
\partial_{\alpha}\mathrm{Im}\chi_{\bf k} \partial_{\beta}\mathrm{Re}\chi_{\bf k} 
- 
\partial_{\beta}\mathrm{Im}\chi_{\bf k} \partial_{\alpha}\mathrm{Re}\chi_{\bf k} 
\right]
\nonumber
\\
&
- \mathrm{Re}\chi_{\bf k}
\left[ 
\partial_{\alpha}\mathrm{Im}\chi_{\bf k} \partial_{\beta}\delta_{\bf k} 
- 
\partial_{\alpha}\delta_{\bf k} \partial_{\beta}\mathrm{Im}\chi_{\bf k} 
\right]
\nonumber
\\
&
+ \mathrm{Im}\chi_{\bf k}
\left[ 
\partial_{\alpha}\mathrm{Re}\chi_{\bf k} \partial_{\beta}\delta_{\bf k} 
- 
\partial_{\alpha}\delta_{\bf k} \partial_{\beta}\mathrm{Re}\chi_{\bf k} 
\right].
\end{align}
With the knowledge of the Berry curvature, we calculate the AHE conductivity, 
\begin{align}\label{AHE}
\sigma_{xy} = \frac{e^2}{\hbar}
\left[ \int_{\mathrm{BZ}}\frac{d^2 k}{(2\pi)^2}\sum_{n } \Omega_{{\bf k};xy;n} 
{\cal F}(\epsilon_{{\bf k};n} ) \right],
\end{align}
where $n$ labels fermion bands, ${\cal F}(\epsilon)$ is the Fermi-Dirac distribution function, and the integration is over the Brillouin zone (BZ). We will be studying what is known as the anitferromagnetic AHE\cite{VonsovskiiTurov1986,Turov1990,TurovShavrov}, i.e. $\sigma_{xy}({\bf M}_{1}+{\bf M}_{2}  = 0;{\bf L}) =  \sigma_{yx}({\bf M}_{1}+{\bf M}_{2}  = 0;-{\bf L})$, when there is no canting of the N\'{e}el order.  

%-----------------------------------------------------------------------------------------------------------------
\section{Genuine antiferromagnets}
%-----------------------------------------------------------------------------------------------------------------
A genuine antiferromagnet is a N\'{e}el ordered collinear antiferromagnet in which crystal symmetry forbids a finite magnetization generated by the N\'{e}el order. Additionally, in a genuine antiferromagnet, the magnetic sublattices are connected by some crystal symmetry operation.
The simplest example is a N\'{e}el order with two magnetic sublattices on a square lattice.
In this case, a combination of translation and time-reversal is the symmetry connecting the two magnetic sublattices. This symmetry keeps the $\mathbf{L}$ vector invariant (each operation, translation and time-reversal, changes the sign of $\mathbf{L}$), while it reverses any magnetization $\mathbf{M}$. Therefore, there is no Dzyaloshinskii's invariant in this system, and finite magnetization cannot be generated by the N\'{e}el order.

%-----------------------------------------------------------------------------------------------------------------
\begin{figure}[h] 

\includegraphics[width=0.6 \columnwidth ]{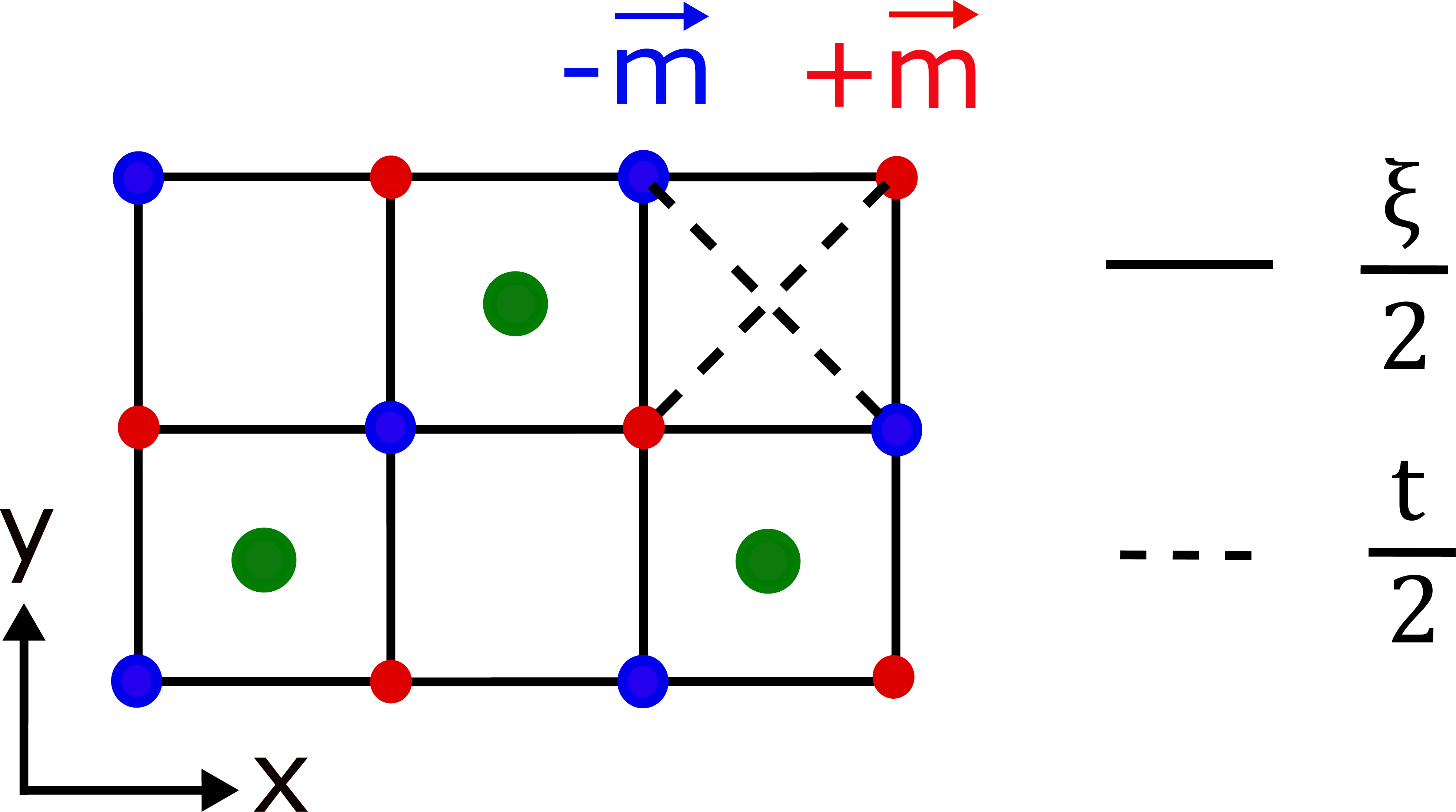} ~~
\includegraphics[width=0.33 \columnwidth ]{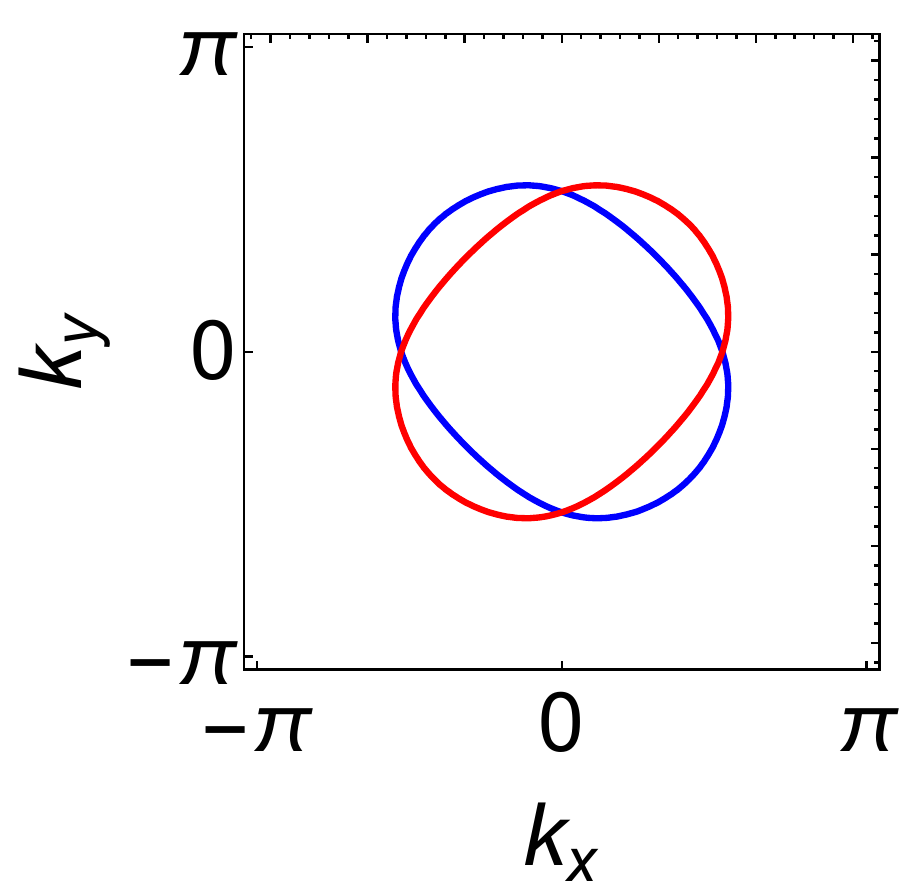}

\protect\caption{
Left: An example of a genuine $d$-wave
N\'{e}el ordered antiferromagnet. The N\'{e}el order is given by the sites with $\pm \mathbf{m}$. The green atom is non-magnetic and can be positioned either in the plane of the lattice or lifted from it. For simplicity, we assume no fermion tunneling through the green atom. Right: Contour plot of the Fermi surfaces of conducting fermions described by Hamiltonian Eq. (\ref{genuine}) for $m=2\xi$ and $t=0.15\xi$, and $\mu = 2.2\xi$ (in units of $\xi$). The N\'{e}el order is in the $z$-direction. The red plot is for spin-up fermions, while the blue is for spin-down.}
\label{fig:fig1}  
\end{figure}
%-----------------------------------------------------------------------------------------------------------------

An antiferromagnet on the square checkerboard lattice (studied for example in \cite{Brekke2023,AgterbergPRB2024,Zyuzin2024}) shown in Fig. (\ref{fig:fig1}) is a hybrid antiferromagnet. Namely, for some directions of the 
N\'{e}el order, the magnetic sublattices are connected with each other, while for other directions they are not connected.  
Non-magnetic green atom eliminates the translation and time-reversal from the symmetries of the crystal. The green atom can be in the plane of the lattice or lifted from it. The magnetic sublattices are connected by a $\frac{\pi}{2}$ rotation and time-reversal for the N\'{e}el vector in $z-$ direction, and for N\'{e}el orders in $x-$ and $y-$ directions there is a combination of reflection in corresponding plane normal to the lattice and time-reversal operation which connects the magnetic sublattices. Therefore, $M_{z} = 0$ for these directions of the N\'{e}el order.

If now the N\'{e}el order is in any other in-plane direction $\cos(\phi) {\bf e}_{x}+\sin(\phi) {\bf e}_{y}$ ($\phi$ is an angle), except for along $x-$ or $y-$ directions, then the two magnetic sublattices will no longer be connected to each other. Technically, the system is a ferrimagnet for these directions of the N\'{e}el order. However, finite $M_{z}$ can't be generated by the N\'{e}el order in this case.
A combination of $\pi$ rotation and time-reversal is the symmetry of the system, which eliminates possible $M_{z}[\cos(\phi) L_{x}+\sin(\phi) L_{y}]$ invariant. This is because $\pi$ rotation and time-reversal restores the N\'{e}el vector, while changes sign of $M_{z}$. In addition to that, if the green atom is in the plane with red and blue sites, a combination of reflection in $x-y$ plane and time-reversal, which is the symmetry of the lattice, which also eliminates possible $M_{z}[\cos(\phi) L_{x}+\sin(\phi) L_{y} ]$ invariant. 

The Hamiltonian of the fermions described by $\Psi = (\Psi_{\mathrm{R};\uparrow},\Psi_{\mathrm{R};\downarrow},\Psi_{\mathrm{B};\uparrow},\Psi_{\mathrm{B};\downarrow})^{\mathrm{T}}$ spinor, where $(...)^{\mathrm{T}}$ is the transposition, is
\begin{align}\label{genuine}
\hat{H}_{\mathrm{genuine}} = \left[ \begin{array}{cc} {\bf m}\cdot {\bm \sigma} - t_{\bf k}  & \xi_{\bf k} \\ 
\xi_{\bf k} &  -{\bf m}\cdot {\bm \sigma} + t_{\bf k} \end{array}\right],
\end{align}
where $\xi_{\bf k} = \xi[\cos(k_x)+\cos(k_{y})]$ and $t_{\bf k} = t\sin(k_{k})\sin(k_{y})$. In deriving $t_{\bf k}$ we assumed that tunneling from red to red along the diagonal is $t^{\mathrm{R}}_{\bf k}=t\cos(k_{x}+k_{y}) = t\cos(k_{x})\cos(k_{y}) -  t\sin(k_{x})\sin(k_{y})$, while from blue to blue it is  $t^{\mathrm{B}}_{\bf k}= t\cos(k_{x})\cos(k_{y}) +  t\sin(k_{x})\sin(k_{y})$. In Eq. (\ref{genuine}) we kept only the second term in $t^{\mathrm{R}}_{\bf k}$ and $t^{\mathrm{B}}_{\bf k}$.
Contour plot of the $d-$wave spin-splitting of fermi surfaces of conducting fermions described by Hamiltonian Eq. (\ref{genuine}) is shown in the right of Fig. (\ref{fig:fig1}). The system shows spin-splitter effect in which spin-up and spin-down polarized currents flow in different directions \cite{Naka2019}. When the Rashba SOC is added for example due to the lifting of the green atom from the plane of the lattice, the system will show $d-$wave Hall effect and linear magnetoconductivity \cite{VorobevZyuzin2024}.

%------------------------------------------------------------------------------------
\section{Ferrimagnet}\label{sectionFerri}
%------------------------------------------------------------------------------------
Ferrimagnets are N\'{e}el ordered antiferromagnets where the two magnetic sublattices, having equal-magnitude spins, are not connected by any symmetry operation. The symmetry between the sublattices is broken by the non-magnetic environment rather than by the difference in the magnitude of antialigned spins. However, the latter are also ferrimagnets.

\begin{figure}[t] 

\includegraphics[width=0.95 \columnwidth ]{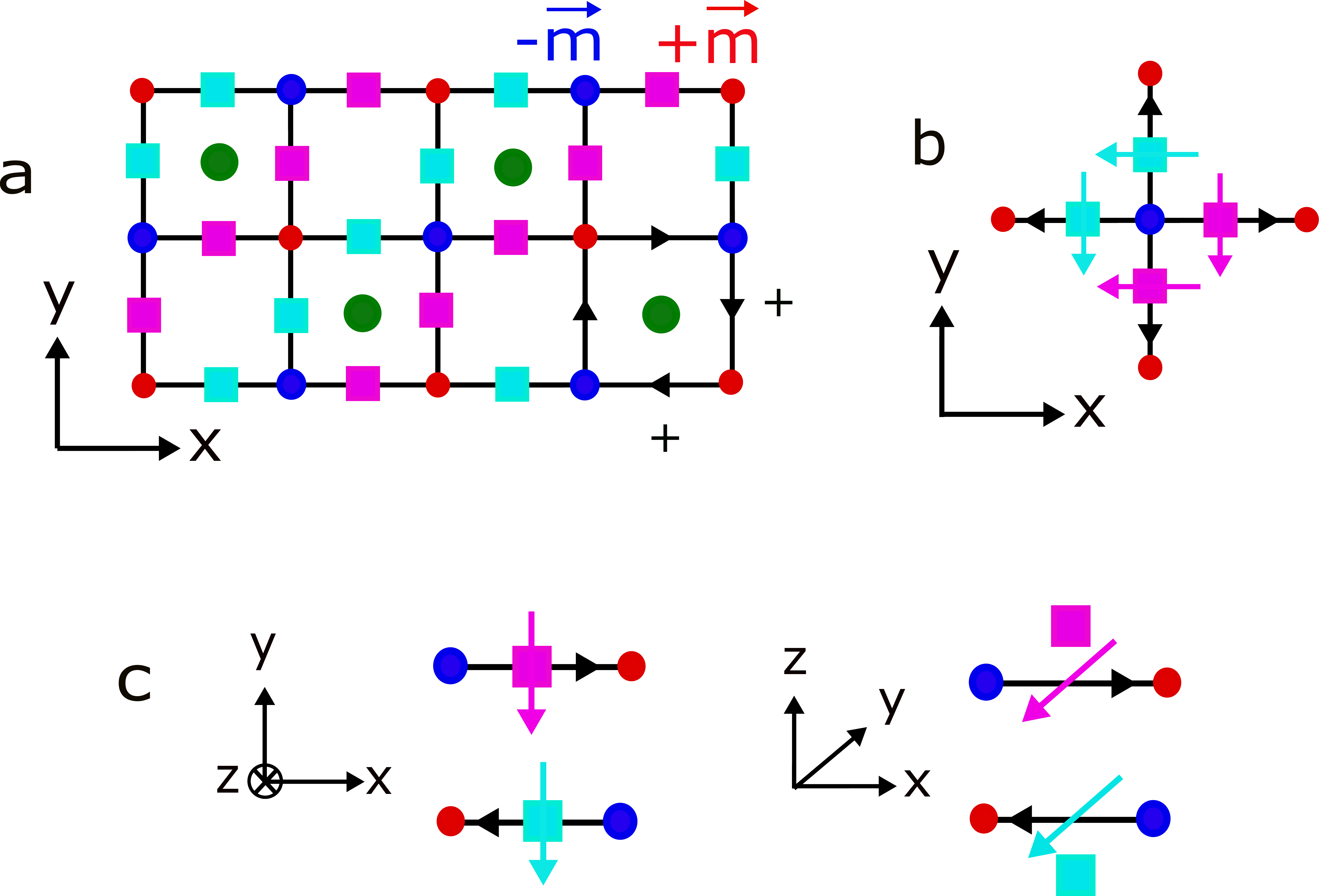} ~~~~

\protect\caption{
(a): Lattice of a N\'{e}el ordered ferrimagnet. (b) Description of the spin-orbit coupling. A square is an atom which is either on the bottom (cyan color) in $z-$direction or on top (purple color) of the link (as shown in (c)). Cyan/purple arrow is the direction of the spin-orbit coupling for a direction of fermion hopping defined by the black arrow. If the direction of the black arrow changes sign, the direction of the cyan/purple arrow will do so as well. There are two Dzyaloshinskii's invariants in the system: $M_{z}L_{z}$ and $M_{z}(L_{x}-L_{y})$.}
\label{fig:fig2}  
\end{figure}

We consider a N\'{e}el ordered system shown in Fig. (\ref{fig:fig2}a). 
The squares represent non-magnetic atoms that are lifted from the $x-y$ plane as shown in Fig. (\ref{fig:fig2}c). 
It can be verified that the magnetic sublattices are not connected by any symmetry operation; hence, the system shown in Fig. (\ref{fig:fig2}) is a ferrimagnet. Let us now figure out which directions of the N\'{e}el order can generate finite $M_{z}$. First consider the N\'{e}el order in $z-$direction. Then a combination of $\frac{\pi}{2}$ rotation about the center of the square plaquette, mirror reflection in the $y-z$ plane which cuts the vertical bond of the square in half, and time-reversal is the symmetry of the lattice, which allows for Dzyaloshinskii's invariant $M_{z}L_{z}$. Now set the N\'{e}el order in the plane of the lattice. In this case the aforementioned combination of symmetry operations is the symmetry of the lattice only when $L_{x}=-L_{y}$. Therefore, Dzyaloshinskii's invariant in this case is $M_{z}(L_{x}-L_{y})$. We, thus, expect AHE in the system to be $\sigma_{xy}\propto \sigma_{z} L_{z} + \sigma_{x-y} (L_{x}-L_{y})$, where $\sigma_{z}$ and $\sigma_{x-y}$ are material dependent coefficients. Let us demonstrate by studying microscopics that the symmetry analysis is correct.

%-----------------------------------------------------------------------------------------------------------------
%-----------------------------------------------------------------------------------------------------------------

As a result of the lifted squares, the mirror symmetry in the $x-y$ plane is broken. The allowed SOC is shown in Fig. (\ref{fig:fig2}b) and (\ref{fig:fig2}c). In addition, in the lower part of the right corner of the lattice shown in Fig. (\ref{fig:fig2}a), there is a SOC of the $d-$wave form created by the green atom. For example, such a SOC was used in Ref. (\onlinecite{Zyuzin2024}).
The basis is $\Psi = (\Psi_{\mathrm{R};\uparrow},\Psi_{\mathrm{R};\downarrow},\Psi_{\mathrm{B};\uparrow},\Psi_{\mathrm{B};\downarrow})^{\mathrm{T}}$ and the Hamiltonian of the system is
\begin{align}\label{ferri}
\hat{H}_{\mathrm{ferri}} = \left[ \begin{array}{cc} {\bf m}\cdot {\bm \sigma} - t_{\bf k}  & \xi_{\bf k} + i{\bm \gamma}_{\bf k}\cdot{\bm \sigma} \\ 
\xi_{\bf k} - i{\bm \gamma}^{*}_{\bf k}\cdot{\bm \sigma} &  -{\bf m}\cdot {\bm \sigma} + t_{\bf k}  \end{array}\right],
\end{align}
where $t_{\bf k} = t\sin(k_{x})\sin(k_{y})$, $\gamma_{\bf k}^{x} = - \gamma\cos(k_{y})$, $\gamma_{\bf k}^{y} = - \gamma\cos(k_{x})$, $\gamma_{\bf k}^{z} = \gamma_{z}\left[ \cos(k_{x}) - \cos(k_{y})\right]$, $\xi_{\bf k} = \xi\left[ \cos(k_{x}) + \cos(k_{y})\right]$, and ${\bf m} =m \left[\cos(\phi)\sin(\theta),\sin(\phi)\sin(\theta),\cos(\theta)\right]$ is in general direction set by $\phi$ and $\theta$ angles. 
We will be interested in studying condcuting fermions only.
For that we found it convenient to rotate $\hat{T}^{-1}{\bf m}\cdot{\bm \sigma}\hat{T} = m\sigma_{z}$, where $m^2=m_{x}^2+m_{y}^2+m_{x}^2$, and then change the basis to conducting and valent fermions
$(\tilde{\Psi}_{\mathrm{R};\uparrow},\tilde{\Psi}_{\mathrm{B};\downarrow},\tilde{\Psi}_{\mathrm{R};\downarrow},\tilde{\Psi}_{\mathrm{B};\uparrow})$, where tilde denotes rotated basis. The Hamiltonian becomes
\begin{align}\label{ferriRotated}
\hat{H}_{\mathrm{ferri}} = \left[ \begin{array}{cc} \hat{H}^{\mathrm{c}} & \hat{C} \\ \hat{C}^{\dag} & \hat{H}^{\mathrm{v}} \end{array}\right],
\end{align}
where 
\begin{align}\label{conductionFerri}
\hat{H}^{\mathrm{c}} = \left[ \begin{array}{cc} m-t_{\bf k} & ia_{{\bf k};12} \\
-ia_{{\bf k};12}^{*}  & m+t_{\bf k} \end{array} \right],  ~\hat{H}^{\mathrm{v}} = \left[ \begin{array}{cc} -m - t_{\bf k} & ia_{{\bf k};12}^{*}  \\
-ia_{{\bf k};12} & -m + t_{\bf k} \end{array} \right],
\end{align}
and $\hat{C} =(\xi_{\bf k} + ia_{\bf k}) \hat{\tilde{\sigma}}_{1}$, where $\hat{\tilde{\sigma}}_{1}$ acts in the rotated basis, and where we have defined
\begin{align}\label{ak12ak}
&
a_{\bf k} = \gamma^{z}_{\bf k} \cos(\theta) + \frac{1}{2} \sin\left( \theta\right)  \left(\gamma^{-}_{\bf k}e^{i\phi}+ \gamma^{+}_{\bf k}e^{-i\phi}\right) 
\\
&
a_{{\bf k};12} = -\gamma^{z}_{\bf k} \sin(\theta) +\gamma^{-}_{\bf k} \cos^2\left( \frac{\theta}{2}\right) e^{i\phi} - \gamma^{+}_{\bf k} \sin^2\left( \frac{\theta}{2}\right)e^{-i\phi}  ,
\end{align}
where $\gamma^{\pm}_{\bf k} = \gamma_{\bf k}^{x} \pm i \gamma_{\bf k}^{y}$, and note that $\gamma^{x/y/z}_{\bf k}$ are real. 

The eignevalue equation for the conducting fermions described by $\Psi_{\mathrm{c}} = (\tilde{\Psi}_{\mathrm{R};\uparrow},\tilde{\Psi}_{\mathrm{B};\downarrow})$, where $\tilde{\Psi}$ is the rotated basis, is obtained to be
\begin{align}
-2m\left[\begin{array}{cc} t_{\bf k} & -ia_{{\bf k};12} \\ ia_{{\bf k};12}^{*}  & -t_{\bf k} \end{array}\right]\Psi_{\mathrm{c}} = E^2\Psi_{\mathrm{c}},
\end{align}
where the phase of $\xi_{\bf k} + ia_{\bf k} = \sqrt{\xi_{\bf k}^2 + a_{\bf k}^2}e^{i\zeta_{\bf k}}$ can also be relevant to the Berry curvature (see Section \ref{sectionRutile}).
\begin{align}
E^2 = (\epsilon + \mu)^2 - m^2 - t_{\bf k}^2 - a_{{\bf k};12}^{*}a_{{\bf k};12} - \xi_{\bf k}^2 -a_{\bf k}^2,
\end{align}
where $\epsilon$ is the eigenvalue.
Spectrum of the two conduction bands is 
\begin{align}\label{spectrumFerri}
\epsilon_{{\bf k};\pm}^{\mathrm{c}} = \sqrt{ \left(m \mp \sqrt{t_{\bf k}^2 + \vert a_{{\bf k};12}\vert^2} \right)^2 + \xi_{\bf k}^2 + a_{\bf k}^2}
\end{align}
The quantity Eq. (\ref{omega}) that defines the Berry curvature is calculated to be
\begin{align}\label{omegaFerri}
\omega_{{\bf k};xy} =& 8t m^2\left[1-\cos^2(k_{x})\cos^2(k_{y})\right]\nonumber
\\
&\times \left[ \gamma^2m_{z} + \gamma_{z}\gamma (m_{y}-m_{x}) \right] ,
\end{align}
and recalling that the N\'{e}el vector is ${\bf L} = 2{\bf m}$, we confirm predictions of the symmetry argument leading to the Dzyaloshinskii's invariant of the system. It should be noted that the Eq. (\ref{omega}) contains all the physical processes that break the symmetries to allow for the magnetic moment in accord with the Dzyaloshinskii's invariant. We observe that SOC, given by $\gamma$, which is due to the breaking of the symmetry of reflection in the $x-y$ plane and time-reversal, enters both expressions in the second line of Eq. (\ref{omegaFerri}).
In addition, the spin splitting given by $t_{\bf k}$ is due to breaking of the translation symmetries in the system. Finally, a combination of $\gamma$'s and $t_{\bf k}$ processes that enters Eq. (\ref{omega}) is the result of breaking the symmetries between the magnetic sublattices by the green atom and colored squares shown in Fig. (\ref{fig:fig2}).

%-----------------------------------------------------------------------------------------------------------------
\begin{figure}[h] 

\includegraphics[width=0.45 \columnwidth ]{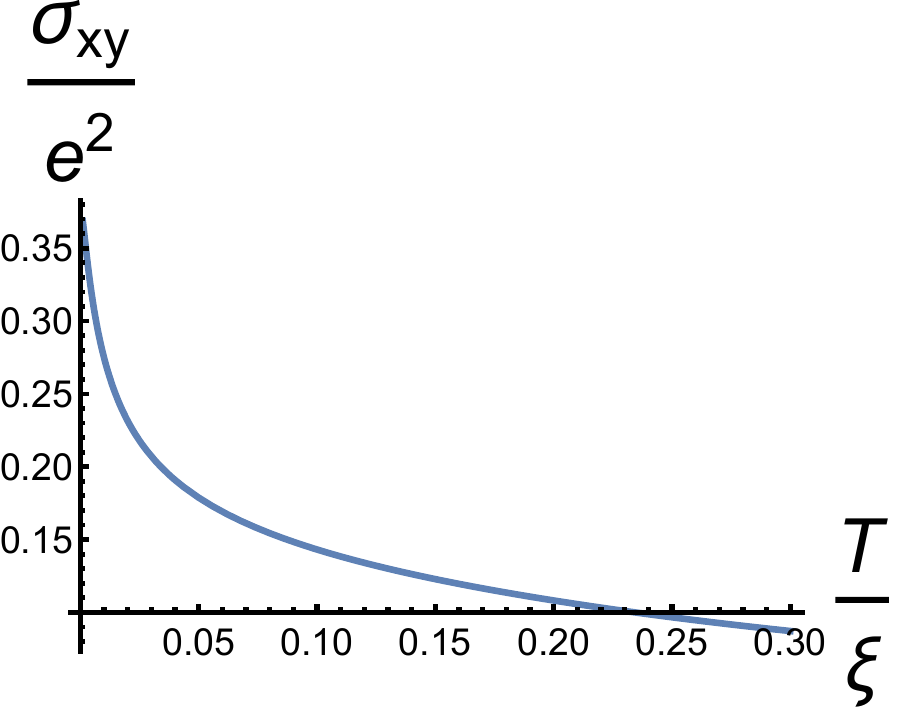} ~~
\includegraphics[width=0.45 \columnwidth ]{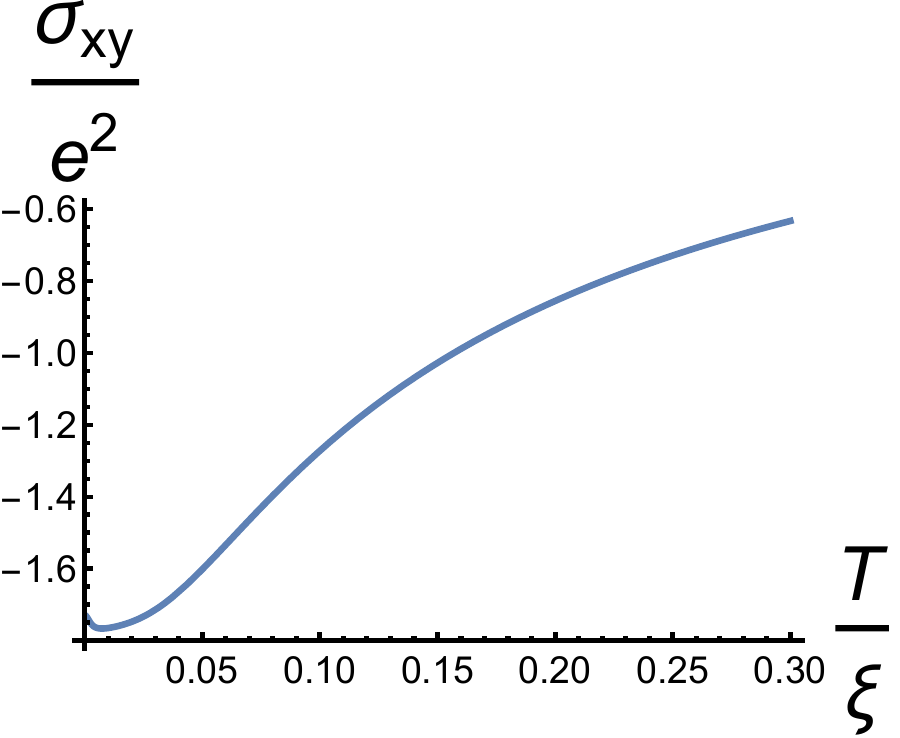} ~~

\protect\caption{
Plot of the AHE conductivity as a function of temperature. Left: ferrimagnetic model Eq. (\ref{ferri}) for $m_{x}=m_{y}=0$. Right: weak ferromagnet model Eq. (\ref{weakMS}) for $m_{x}=m_{z}=0$. In both plots $m=2\xi$, $t=0.15\xi$, $\gamma=\eta_{1}=\eta_{2}=0.1\xi$, and Fermi level was chosen $\mu = 2\xi$. It was assumed that $h=2\pi \hbar \equiv 1$.}
\label{fig:fig3}  
\end{figure}
%-----------------------------------------------------------------------------------------------------------------

%----------------------------------------------------------------------------------
\section{Weak ferromagnet}\label{sectionWeak}
%----------------------------------------------------------------------------------

Weak ferromagnetism of antiferromagnets has been first proposed by I.E. Dzyaloshinskii in 1958 \cite{Dzyaloshinskii1958} as an explanation of experiments by A.S. Borovik-Romanov \cite{BorovikRomanov1957}. 
A weak ferromagnetic is a N\'{e}el ordered antiferromagnet in which crystal symmetry allows for the existence of a finite magnetic moment. This magnetic moment may be either canting of the N\'{e}el order\cite{BorovikRomanov1957,Dzyaloshinskii1958}, or be due to the orbital magnetization of conducting fermions. Below we will be interested in the latter. In weak ferromagnets, contrary to ferrimagnets, magnetic sublattices of the N\'{e}el order are connected to each other by a symmetry operation.\cite{CommentAlter}.

\begin{figure}[h] 

\includegraphics[width=0.9 \columnwidth ]{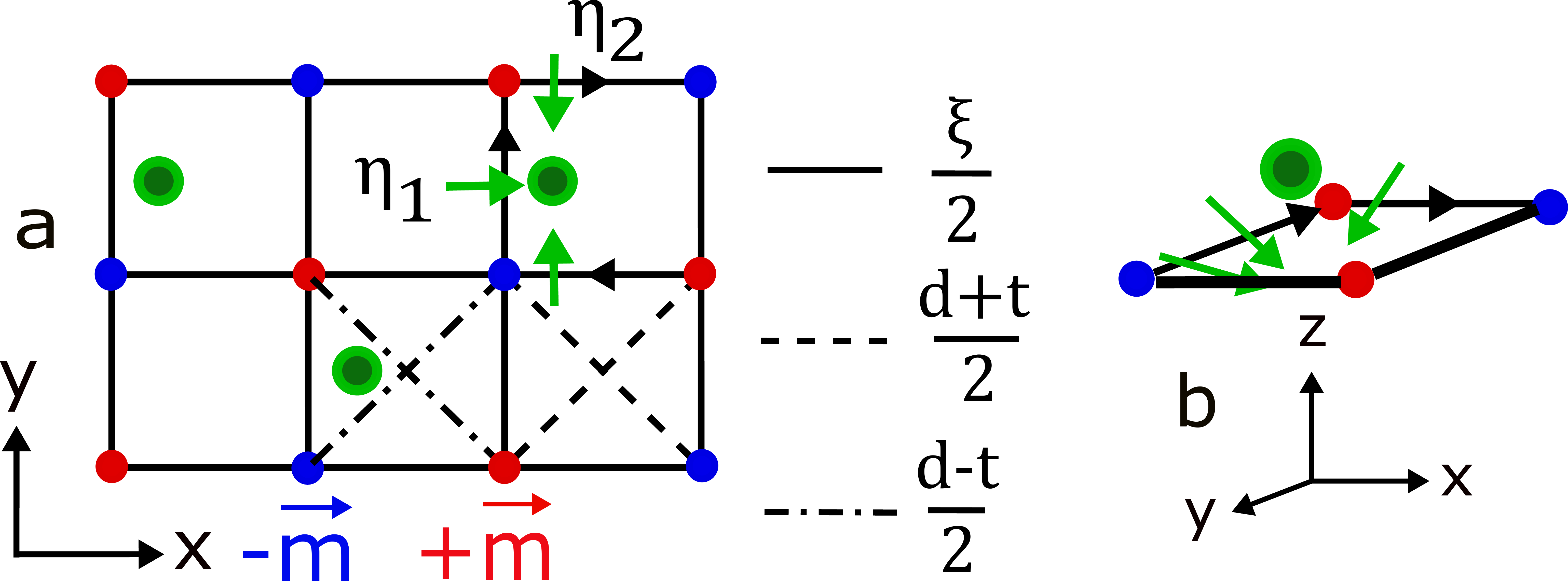} ~~~~

\protect\caption{
(a) A model of mirror-symmetric weak ferromagnet. A combination of a mirror reflection in the $x-z$ plane (crossing the center of the vertical link) and time-reversal operations is the symmetry which connects the two sublattices. (b) The green atom is lifted from the plane of the lattice. This is needed to eliminate a combination of reflection in the $x-y$ plane and time-reversal from the symmetries of the system. As a result, SOC acquires in-plane components shown by the green arrows. Thus, Dzyaloshinskii's invariant of weak Dzyaloshinskii's ferromagnetism is $M_{z}L_{y}$ in this model.}
\label{fig:fig4}  
\end{figure}

We aim to construct a simple theoretical model of a weak ferromagnet. In Ref. (\onlinecite{Zyuzin2025a}) a model of genuine mirror-symmetric antiferromagnet has been proposed. 
In Fig. (\ref{fig:fig4}) a generalization of the model of Ref. (\onlinecite{Zyuzin2025a}) of the genuine antiferromagnet to the case of the weak ferromagnet is shown. 
Let us determine the non-zero Dzyaloshinskii's invariant in this system. If the N\'{e}el order is in the $x$- or $z$-direction, the symmetry connecting the two magnetic sublattices is a combination of reflection in the $x$-$z$ plane (crossing the vertical link center) and time-reversal. This combination ensures that $L_{x}$ or $L_{z}$ do not change sign, while $M_{z}$ does.
Therefore $M_{z}L_{x/z}$ isn't the Dzyaloshinskii's invariant of the system.
When the N\'{e}el order is in $y-$ direction, the symmetry of the system which connects the two magnetic sublattices is a combination of reflection in the $x-z$ plane which crosses the vertical link in the center, and time-reversal operation. Reflection reverses both $L_{y}$ and $M_{z}$. Then, both $L_{y}$ and $M_{z}$ change sign under the time-reversal operation. 
Therefore, this symmetry allows for $M_{z}L_{y}$ to be the Dzyaloshinskii's invariant of the system.
However, if the green atom is placed strictly in the plane of the lattice, then a combination of reflection in the $x-y$ plane and time-reversal operation is the symmetry of the lattice, which keeps $M_{z}$ intact but reverses the sign of $L_{y}$. To allow for the $M_{z}L_{y}$ Dzyaloshinskii's invariant, we must break this symmetry by lifting the green atom from the lattice plane, as shown in Fig.~\ref{fig:fig4}. We thus expect the AHE to be $\sigma_{xy}\propto \sigma_{y} L_{y}$, where $\sigma_{y}$ is a material-dependent coefficient.

%-----------------------------------------------------------------------------------------------------------------
%-----------------------------------------------------------------------------------------------------------------

We now demonstrate that the symmetry argument is consistent with microscopic calculations of the Berry curvature.
Fermion tunneling between red and blue sites along the diagonal (dashed and dashed-dotted lines in Fig. (\ref{fig:fig4}) is 
\begin{align}
t_{\bf k}^{\mathrm{R}/\mathrm{B}}=
&
\frac{d \pm t}{2}\cos(k_{x} \pm k_{y}) + \frac{d \mp t}{2}\cos(k_{x} \mp k_{y}) \nonumber
\\
&= d\cos(k_{x})\cos(k_{y}) \mp t\sin(k_{x})\sin(k_{y}).
\end{align}
We omit the $d\cos(k_{x})\cos(k_{y})$ term in the following calculation, since it is the same for both sublattices, and it doesn't affect the orbital magnetization of fermions. It will be restored in the calculation of the AHE.
In the basis of $\hat{\Psi} = (\Psi_{\mathrm{R}\uparrow},\Psi_{\mathrm{B}\downarrow},\Psi_{\mathrm{B}\uparrow},\Psi_{\mathrm{R}\downarrow})^{\mathrm{T}}$,
the Hamiltonian of the model, containing all the necessary ingredients for the non-zero orbital magnetization is
\begin{align}\label{weakMS}
\hat{H}_{\mathrm{weak}} = 
\left[\begin{array}{cc}
{\bf m}\cdot{\bm \sigma}  - t_{{\bf k}}  & \xi_{\bf k} + i{\bm \gamma}_{\bf k}\cdot{\bm \sigma}
\\
\xi_{\bf k} - i{\bm \gamma}^{*}_{\bf k}\cdot{\bm \sigma} & -{\bf m}\cdot{\bm \sigma}  + t_{{\bf k}} 
\end{array}\right],
\end{align}
where $\xi_{\bf k} = \xi [\cos(k_{x})+\cos(k_{y})]$ and $t_{\bf k} = t \sin(k_{x})\sin(k_{y})$. The structure of the spin-orbit coupling created by the green atom is shown in Fig. (\ref{fig:fig4}b), and is given by
$\gamma^{x}_{\bf k} = \eta_{1}e^{ik_{y}}$, $\gamma^{z}_{\bf k} = - \eta_{2}\cos(k_{x}) + \eta_{4}e^{ik_{y}}$, and $\gamma^{y}_{\bf k} = i\eta_{3}\sin(k_{x})$. 
The $i\eta_{1}\sin(k_{y})$ and $\gamma^{y}$ components are standard Rashba SOC. The part of $\gamma^{x}_{\bf k}$ with $\eta_{1}\cos(k_{y})$ is due to the lowering of the symmetry by the position of the green atom. Indeed, such a position eliminates all symmetries of the genuine antiferromagnet discussed above in Eq. (\ref{genuine}).

It is again convenient to rotate the spin basis of Eq. (\ref{weakMS}) as it was done for ferrimagnet system Eq. (\ref{ferri}). Furthermore, it is useful to rearrange the basis to conduction and valence bands as $\hat{\tilde{\Psi}} = (\tilde{\Psi}_{\mathrm{R}\uparrow},\tilde{\Psi}_{\mathrm{B}\downarrow},\tilde{\Psi}_{\mathrm{R}\downarrow},\tilde{\Psi}_{\mathrm{B}\uparrow})^{\mathrm{T}}$, where $\tilde{\Psi}$ is the rotated basis. We define $\xi_{\bf k} + ia_{\bf k} = \sqrt{(\xi_{\bf k}-\mathrm{Im}a_{\bf k})^2 + (\mathrm{Re}a_{\bf k})^2}e^{i\zeta_{1;\bf k}}$ and $\xi_{\bf k} + ia^{*}_{\bf k} = \sqrt{(\xi_{\bf k}+\mathrm{Im}a_{\bf k})^2 + (\mathrm{Re}a_{\bf k})^2}e^{i\zeta_{2;\bf k}}$. These phases can be relevant to the Berry curvature (see Section \ref{sectionRutile}). In this new basis the Hamiltonian is
\begin{align}
\hat{H}_{\mathrm{weak}} = \left[ \begin{array}{cc} \hat{H}^{\mathrm{c}} & \hat{C} \\ \hat{C}^{\dag} & \hat{H}^{\mathrm{v}} \end{array}\right],
\end{align}
where the Hamiltonian of the conduction band is
\begin{align}\label{conductionWeak}
\hat{H}^{\mathrm{c}} = \left[ \begin{array}{cc} m-t_{\bf k} & ia_{{\bf k};12} \\
-ia_{{\bf k};12}^{*}  & m+t_{\bf k} \end{array} \right],  \hat{H}^{\mathrm{v}} = \left[ \begin{array}{cc} -m - t_{\bf k} & ia_{{\bf k};21}  \\
-ia^{*}_{{\bf k};21} & -m + t_{\bf k} \end{array} \right],
\end{align}
and the expression for $\hat{C}$ is
\begin{align}
 \hat{C} =  \left[ \begin{array}{cc} 0 & \xi_{\bf k} + ia_{\bf k}  \\  \xi_{\bf k} + ia^{*}_{\bf k} & 0 \end{array}\right].
\end{align}
Quantities $a_{\bf k}$ and $a_{{\bf k};12}$ are defined in Eq. (\ref{ak12ak}), in which $\gamma^{x/y/z}_{\bf k}$ are given after Eq. (\ref{weakMS}). In addition, we have defined
\begin{align}
a_{{\bf k};21} = -\gamma^{z}_{\bf k} \sin(\theta) +\gamma^{+}_{\bf k} \cos^2\left( \frac{\theta}{2}\right) e^{-i\phi} - \gamma^{-}_{\bf k} \sin^2\left( \frac{\theta}{2}\right)e^{i\phi}.
\end{align}
The equation defining conduction band described by $\hat{\Psi}_{\mathrm{c}} = (\tilde{\Psi}_{\mathrm{R}\uparrow},\tilde{\Psi}_{\mathrm{B}\downarrow})$ spinor is 
\begin{align}
(\hat{H}^{\mathrm{c}} - E)\hat{\Psi}_{\mathrm{c}} - \hat{C} (\hat{H}^{\mathrm{v}} - E)^{-1} \hat{C}^{\dag}\hat{\Psi}_{\mathrm{c}} = 0,
\end{align}
and for the purposes of obtaining analytical expressions of the Berry curvature of the conducting fermions, it is safe to analyze only the Hamiltonian Eq. (\ref{conductionWeak}). Other terms coming from $\hat{H}^{\mathrm{v}}$ are small in $\frac{1}{m}$. Analtyical expression for the Berry curvature for general direction of the N\'{e}el order is complicated, but we can check different special cases.
We set $\eta_{3}=\eta_{4}=0$ and $m_{x}=m_{z} = 0$, get $a_{\bf k} = 0$ and $a_{{\bf k};12} = \eta_{2}\cos(k_{x}) + i\frac{m_{y}}{m}\eta_{1}e^{ik_{y}}$, and obtain
\begin{align} \nonumber
\omega_{{\bf k};xy} &= t\eta_{1}\eta_{2}\frac{m_{y}}{m}\left[ 1 - \cos^2(k_{x})\cos^2(k_{y})\right]
\\
&
+t\eta_{1}^2\left( \frac{m_{y}}{m} \right)^2\sin(k_{y})\cos(k_{x}), \label{omegaWeak}
\end{align}
where second term will vanish upon integration over the BZ.
Let us now demonstrate that other directions of the N\'{e}el order will result in zero AHE. We set $m_{y}=m_{z} = 0$ and pick spin-orbit coupling $\eta_{1}=\eta_{2}=0$, then  $a_{\bf k} = 0$ and $a_{{\bf k};12} = -\eta_{4}e^{ik_{y}} + \frac{m_{x}}{m} \eta_{3}\sin(k_{x})$. The curvature is calculated to be
\begin{align}
\omega_{{\bf k};xy}= - \frac{m_{x}}{4m}t\eta_{3}\eta_{4}\sin(2k_{x})\sin(2k_{y}) + t\eta_{4}^2 \sin(k_{y})\cos(k_{x}),
\end{align}
which would be integrated to zero. Other combinations with other $\eta$'s and $m_{\alpha}$ will be integrated to zero in a similar way. All in all, the symmetry argument of the existence of finite magnetization in the studied system is consistent with the microscopic calculation of the Berry curvature. Only the $m_{y}\neq 0$ and lifting of the green atom from the plane of the lattice, characterized by $\eta_{1}$ parameter, are important in obtaining finite magnetic moment in the system. In addition, the asymmetry of the SOC $\eta_{1}$ which is due to the shifted in-plane position of the green atom from the center of the square plaquette, is crucial. Indeed, if the green atom was in the center of the square and lifted from the plane, the corresponding SOC would be $\propto i\eta_{1}\sin(k_{x})$ which is not in favor of the magnetic moment. In addition, a combination of $t$ and $\eta_{2}$ eliminates symmetries of the genuine antiferromagnets. We plot the AHE conductivity of the model for $m_{y}\neq 0$ as a function of temperature in Fig. (\ref{fig:fig3}). In deriving the Berry curvature we used eigenfunctions numerically derived from Eq. (\ref{weakMS}) rather than from reduced Hamiltonian Eq. (\ref{conductionWeak}).

%----------------------------------------------------------------------------------
\section{Weak ferromagnet on a rutile lattice}\label{sectionRutile}
%----------------------------------------------------------------------------------
Antiferromagnet on a rutile lattice is a classical example of a weak ferromagnet \cite{Dzyaloshinskii1958}.
A combination of $\frac{\pi}{2}$ rotation, translation and time-reversal forbids N\'{e}el order in ${\bf e}_{z}$ from developing any magnetization in the system.
Dzyaloshinskii's invariants are $M_{x}L_{y}$ and $M_{y}L_{x}$, and a pseudovector ${\bf M}$ is allowed in this system. Again, this pseudovector may be the Dzyaloshinskii's N\'{e}el order canting, or the orbital magnetic moments of fermions or magnons. CoF$_2$, MnF$_2$, FeF$_2$\cite{BorovikRomanov1960,BorovikRomanov1973} are the examples of a weak ferromagnet on the rutile lattice. The direction of the N\'{e}el order in these materials is along the $z-$axis, and there is no pseudovector in them. These materials are insulating. RuO$_2$ is a metallic antiferromagnet on a rutile lattice with N\'{e}el order in $z-$direction. Insulating NiF$_2$ is another antiferromagnet on the rutile lattice believed to have the N\'{e}el vector in the $x-y$ plane\cite{BorovikRomanov1973}. We nevertheless set the N\'{e}el order in the $x-y$ plane and assume that the spins on the two sublattices are collinear to each other. We wish to understand whether the pseudovector ${\bf M}$ develops in the orbital magnetization of fermions that interact with the N\'{e}el order.

  %-----------------------------------------------------------------------------------------------------------------
\begin{figure}[h] 

\includegraphics[width=0.3 \columnwidth ]{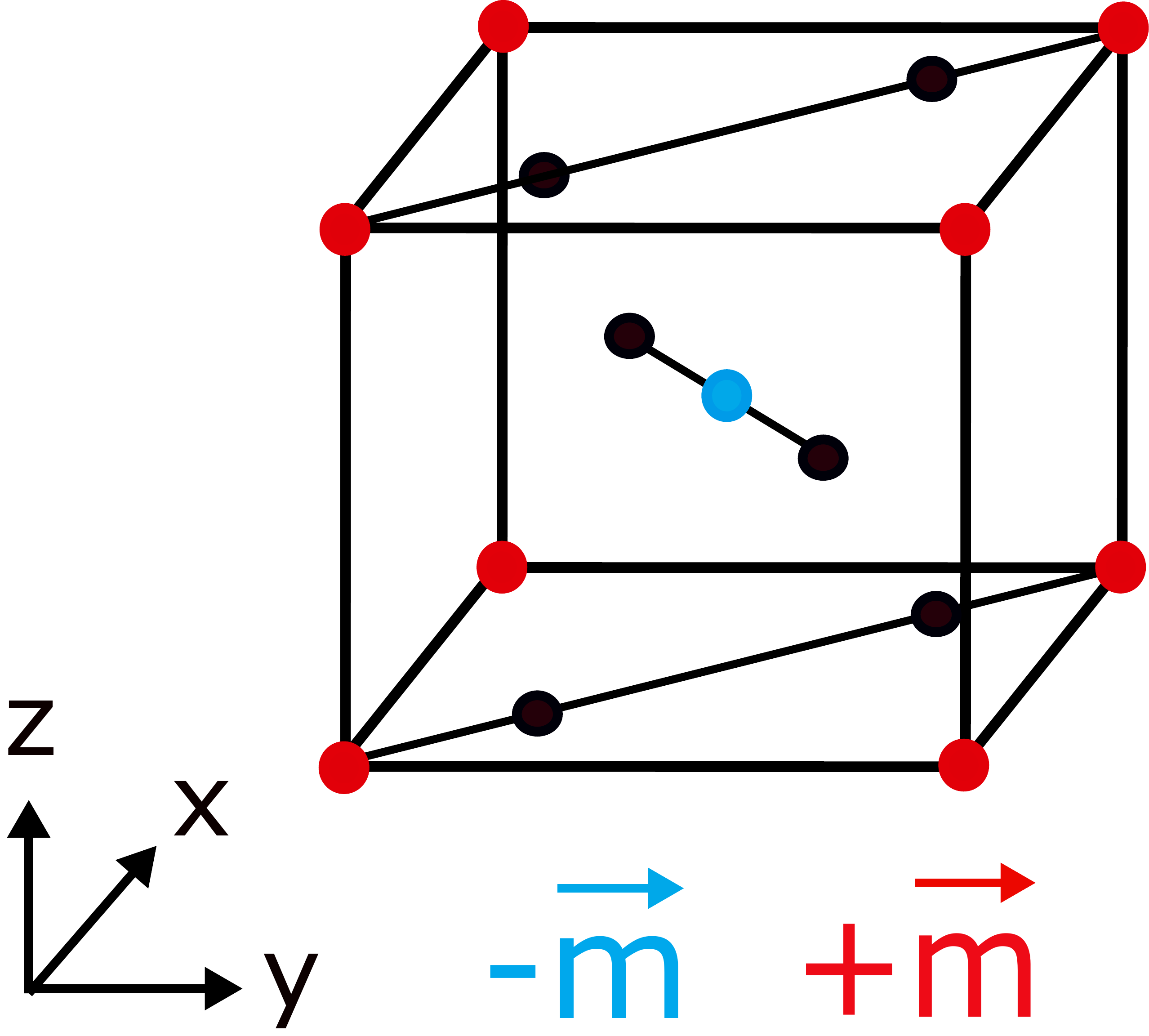} ~~~~

\protect\caption{
N\'{e}el order on the rutile lattice. Black sites are non-magnetic atoms. Red and blue sites correspond to magnetic sites with $\pm{\bf m}$ magnetization. Black sites create the only spin-orbit coupling in the system. Fermions reside on red and blue sites.}
\label{fig:fig5}  
\end{figure}
%-----------------------------------------------------------------------------------------------------------------
Hamiltonian of conducting fermions on the rutile lattice have the structure of Eq. (\ref{ferri}) with the following elements, 
 $\xi_{\bf k} = \xi \cos(k_{x})\cos(k_{y})\cos(k_{z})$, $\gamma^{x}_{\bf k} = -\gamma \sin(k_{z})\sin(k_{x})\cos(k_{y})$, $\gamma^{y}_{\bf k} = \gamma \sin(k_{z})\cos(k_{x})\sin(k_{y})$, and $t_{\bf k} = t\sin(2k_{x})\sin(2k_{y})$. We set the N\'{e}el order to be in ${\bf e}_{x}$ direction. The approach applied in Section \ref{sectionFerri} is relevant to the system at hand. After the rotation is performed, we get $a_{\bf k} = \gamma_{\bf k}^{x}$, $a_{{\bf k};12} = -i \gamma_{\bf k}^{y}$. It is straightforward to show that $a_{{\bf k};12}$ and $t_{\bf k}$ result in zero Berry curvature. We set them to zero and study the phase of $\hat{C}$ matrix in Eq. (\ref{ferriRotated}) by reducing the Hamiltonian into two equivalent blocks,
 \begin{align}\label{rutileApprox1}
\hat{H}_{\mathrm{rutile}} = \left[
\begin{array}{cc} 
 m_{x} & \xi_{\bf k} + ia_{\bf k} \\
 \xi_{\bf k} - ia_{\bf k} & -m_{x} \\
\end{array} \right],
\end{align}
written in the basis of $(\tilde{\Psi}_{\mathrm{R};\uparrow},\tilde{\Psi}_{\mathrm{B};\uparrow})^{\mathrm{T}}$ and $(\tilde{\Psi}_{\mathrm{B};\downarrow},\tilde{\Psi}_{\mathrm{R};\downarrow})^{\mathrm{T}}$. The Berry curvature corresponding to magnetic moment in ${\bf e}_{y}$ direction is
\begin{align}\label{BerryRutile}
\Omega_{{\bf k};xz}^{\pm} = \pm  \frac{m_{x} \xi \gamma \cos^{2}(k_{y})\left[ \sin^{2}(k_{x}) - \sin^2(k_{z})\right]}{2\left[m_{x}^2+ \xi_{\bf k}^2 +a_{\bf k}^2 \right]^{\frac{3}{2}}},
\end{align}
and it may appear that it will also be integrated to zero. However, the AHE is non-zero by the virtue of $a_{{\bf k};12}$ and $t_{\bf k}$ in the spectrum Eq. (\ref{spectrumFerri}). These terms are not important in the Berry curvature but must be restored in the spectrum since they break the symmetry in $x$ and $z$ directions. The $\pm$ sign in Eq. (\ref{BerryRutile}) corresponds to conduction/valence bands. The signs of the Berry curvature for the conduction bands are of the same sign. This is in contrast to the weak ferromagnet discussed in the previous section.
The magnitude of AHE conductivity is tiny for realistic parameters. We conclude that the provided calculation of the Berry curvature is consistent with the $M_{y}L_{x}$ Dzyaloshinskii's invariant. Presence of the second invariant $M_{x}L_{y}$ in the system can be proven in the same way. Overall, the AHE is ${\bf j}^{\mathrm{AHE}} = \sigma_{xz}L_{x} [{\bf e}_{y}\times {\bf E}] + \sigma_{yz}L_{y}[{\bf e}_{x}\times {\bf E}] $.

%-----------------------------------------------------------------
\section{Conclusions}
%-----------------------------------------------------------------
In this paper we have constructed two-dimensional theoretical minimal models of N\'{e}el ordered collinear metallic antiferromagnets that show the AHE. We emphasize that we were interested in the case when the AHE is due to the N\'{e}el vector ${\bf L}$ and there is no canting of the N\'{e}el order in the system.  
We have demonstrated that only collinear ferrimagnets and collinear weak ferromagnets show the AHE. \cite{CommentAlter}
Ferrimagnets are the N\'{e}el ordered antiferromagnets, i.e. with equal in magnitude and antialigned spins, which have no symmetry connecting the magnetic sublattices. Weak ferromagnets, on the other hand, are N\'{e}el ordered antiferromagnets which have a symmetry that connects the magnetic sublattices. We have analyzed and obtained Dzyaloshinskii's invariants of the existence of finite magnetization in N\'{e}el ordered antiferromagnets for our proposed theoretical models. 
Microscopic calculations of the Berry curvature for our theoretical models confirmed the structure of the Dzyaloshinskii's invariants. We have identified two main mechanisms of the AHE in collinear ferrimagnets and weak ferromagnets.
The first one is the interplay of momentum-dependent spin-splitting of conducting fermions due to interaction with the N\'{e}el order, and the crystal symmetry-allowed spin-orbit coupling. 
All of the ingredients for a non-zero AHE originate from the broken symmetries that permit the Dzyaloshinskii's invariants in the system.

\section{Acknowledgements}
We thank I.V. Solovyev and G.E. Volovik for discussions.
VPG thanks Landau Institute's 2025 Summer School of Theoretical Physics.
VAZ is grateful to Pirinem School of Theoretical Physics for hospitality during the Summer of 2025. 
This work is supported by FFWR-2024-0016.

\end{document}